\documentclass{optica-article}

\journal{opticajournal} 

\articletype{Research Article}

\usepackage{
    bm,
    lineno,
    textcomp, 
    float,
    mathrsfs,
    booktabs,
    hhline,
    tabularray,
    upgreek
    }

\newcommand{\vect}[1]{\mathbf{#1}}
\newcommand{\vectrho}{\bm{\rho}}
\newcommand{\appenref}[1]{Appendix A}
\newcommand{\secref}[1]{Sec.~\ref{#1}}
\newcommand{\eqnref}[1]{Eq.~(\ref{#1})}
\newcommand{\figref}[1]{Fig.~\ref{#1}}

\newcommand{\maxwell}[0]{\text{s.t. }\ \nabla\times\frac{1}{\mu}\nabla\times\vect{E}-\omega_m^2\vect{\varepsilon}(\vectrho)\vect{E}=\ -j\omega_m\vect{J}}

\begin{document}

\title{Inverse Design for Nonplanar Conformal Layers}

\author{Arjun Khurana, Michael J. Probst, and Stephen E. Ralph\authormark{*}}

\address{School of Electrical and Computer Engineering, Georgia Institute of Technology, Atlanta, Georgia\\}

\email{\authormark{*}stephen.ralph@ece.gatech.edu}


\begin{abstract*} 
Conformal layering, where a deposited layer follows the nonplanar topography beneath it, arises from both deliberate passivation overlays and imperfect planarization between depositions. However, inverse design methods typically model planar, vertically extruded layers, resulting in geometric discrepancies between the modeled and fabricated structure. To address these discrepancies, we present differentiable mapping functions that incorporate both uniform conformal overlays and partially conformal interlayer dips into a standard inverse design paradigm, requiring no additional design parameters over planar structures. Using these mappings, we design (1) mode-selective GaAs grating reflectors with uniform conformal SiN passivation for high-aspect-ratio VCSELs, reaching differential modal reflectivities of 33.1\% (single-etch) and 63.4\% (dual-etch), and (2) a dual-layer silicon grating coupler whose upper layer dips up to 40~nm into the lower layer, reaching $-0.75$~dB coupling efficiency. Modeling conformal layering during optimization, rather than treating it as post-hoc fabrication error, permits direct exploitation of the conformal layering, yielding inherently fabricable and performant designs.
\end{abstract*}

\section{Introduction}

\begin{figure}
    \centering
    \includegraphics{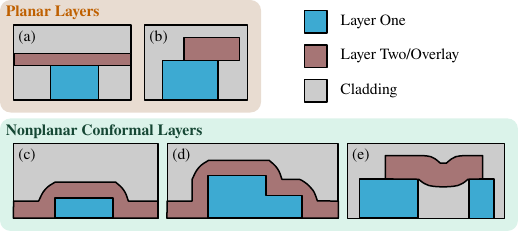}
    \caption{Cross-sectional schematics of planar and nonplanar multilayer photonic structures. Blue regions denote the first device layer, red regions the second layer or overlay, and gray regions the cladding. (a) After the first layer is etched, backfilled with cladding, and planarized, a uniform-thickness overlay is deposited on the flat surface. (b) Both layers are independently deposited, etched, backfilled, and planarized, producing two patterned layers with perfectly vertical sidewalls and flat interfaces. Planar layers are readily handled by conventional density-based TO \cite{probst2024fabrication}. (c, \secref{sec:vcsel}) The second layer is deposited directly after the first layer is etched, without backfilling or planarization, so it conforms to the underlying topography and forms a fully conformal overlay with rounded corners. (d, \secref{sec:vcsel}) Multiple etch steps at different depths give the first layer a stepped, pyramid-like profile, which is then covered by a uniform conformal overlay. (e, \secref{sec:multilayer}) Imperfect planarization after backfilling leaves the cladding recessed in the etched gaps of the first layer. The subsequently deposited and patterned second layer follows this uneven surface and dips down into the gaps.}
    \label{fig:all_layers}
\end{figure}

As photonic integrated circuits scale toward denser, more heterogeneous systems, devices increasingly rely on multiple deposited layers: passivation layers protect etched semiconductor surfaces, and multilayer stacks (\figref{fig:all_layers}a,b) provide the vertical asymmetry and local index control that enable high-performance fiber-to-chip coupling \cite{kohli2023c,hammond2022high,10799716}, optical filtering \cite{zhu2025reconfigurable}, and 3D heterogeneous integration \cite{khurana2025inverse}. In many fabrication contexts, however, each deposited layer takes the shape of the surface beneath it. In particular, when a layer is deposited directly onto an etched structure, or when planarization between subsequent depositions is imperfect, the upper layer conforms to the nonplanar topography below. This conformal layering may arise either as a deliberate design choice \cite{liu2008conformal,xu2020broadband} or as an undesirable fabrication byproduct \cite{probst2024fabrication,Lee2024PolarizationSplitterRotator}, compounding other multilayer challenges such as interlayer misalignment \cite{probst2024fabrication}, increased process complexity \cite{probst2025integrated}, and layer thickness variation. Inverse design methods typically represent each layer as a planar, vertically extruded pattern, which is incapable of modeling this conformal layering. In this paper, we develop differentiable mapping functions that incorporate both uniform conformal overlays and partially conformal interlayer dips directly into the inverse design pipeline. Because optical gratings are fundamental building blocks for routing and controlling light \cite{ChengEtAl2020GratingCouplers,hammond2022multi,HuangBarz2025VerticalCoupler} and are highly sensitive to layer structure, we demonstrate the method with two grating case studies: surface-relief grating reflectors with conformal dielectric passivation for mode selectivity in vertical-cavity surface-emitting lasers (VCSELs), and a dual-layer silicon grating coupler with an imperfectly planarized lower layer.

Topology optimization (TO) \cite{ChristiansenSigmund2021,BallewEtAl2023Discrete,LupoiuEtAl2025MultiAgent,SchulEtAl2026Review,hammond_constraints,hammond2022high,JensenSigmund2011,Molesky2018} is a flexible form of inverse design that parameterizes the device permittivity into hundreds or thousands of voxels, and optimizes these voxels as unique degrees of freedom. The optimization relies on gradient-based techniques to maximize device performance, producing freeform devices with compact footprints, unintuitive device geometries, and high performance. TO's applicability across such a broad range of design problems stems from its ability to directly incorporate physical properties of the device — such as minimum feature size enforcement, process variation, and robustness — into the optimization itself. Realizing this benefit, however, requires designing the mathematical transforms and optimization techniques — such as constraints on the design region for feature size, worst-case robustness formulations, and sophisticated projections — that translate these \textit{physical} realities of the optical device into a form compatible with the \textit{mathematical} nature of the optimization.

\subsection{Case Study: Mode-Selective VCSEL Gratings}
Optical gratings are widely used in photonic design for the unique property of periodic modulation of optical index, enabling different behavior per wavelength, polarization, and k-vector. An illustrative example of this is for polarization control in VCSELs: several works have demonstrated integration of high contrast grating structures for polarization selectivity (and therefore lower emission linewidth) for conventional round-mesa VCSELs \cite{change2009grating}. For high power applications, however, round VCSELs are suboptimal. Simply scaling the area of a round VCSEL yields many high-order modes, which lase at different wavelengths and produce a broad emission spectrum. Thus, when high power and single mode operation are required, low-dimensional high-aspect-ratio rectangular apertures demonstrate a significant advantage over large round VCSELs \cite{Trumpf26vcsel}. Within these structures, many modes exist along the long axis of the rectangular or racetrack aperture, and typically only a single mode exists along the short axis. Thus, linear gratings are highly effective at mode control for these structures. Additionally, non-periodic or quasi-periodic gratings, such as segmented gratings, can be used to create a subwavelength-structured optical phase mask which preferentially selects one or more particular modes while also conditioning the near- and far-fields \cite{Trumpf26vcsel}. 

Typical VCSELs include an undoped semiconductor cap layer, grown as the final epitaxial layer, and a dielectric passivation layer, deposited after the mesa etch, which respectively set the reflection phase for the top DBR and protect the etched mesa from further oxidation. Etching gratings into the dielectric passivation layer has been shown to have some mode selective impact \cite{Khurana2026VCSEL}; however, the index contrast between typical passivation dielectric and air is relatively poor. We therefore consider a fabrication workflow that first etches the semiconductor cap layer and then deposits the passivation nitride. Consequently, the thickness of the passivation is a critical consideration: a thick, planarized passivation layer simplifies grating reflector design but reduces the index contrast between grating teeth and requires additional process steps. Instead, we consider a relatively thin, conformal passivation layer, maintaining a high effective index contrast for the grating teeth and eliminating the need for post-fab planarization (\figref{fig:all_layers}c,d). Incorporating this conformal layer into the grating reflector simulation and optimization therefore introduces an important additional modeling consideration into the design pipeline.

\subsection{Case Study: Multilayer Grating Couplers}
Within the context of grating couplers, multiple layers extend the basic single layer grating design, allowing finer control over the local effective index profile, introducing asymmetry into the design, and therefore enabling higher coupling efficiencies, broader bandwidths, and reduced back-reflection \cite{hammond2022high,probst2024topology}. However, multilayer structures incur increased fabrication complexity, since each additional layer requires precise thickness control, alignment, and often multiple lithography and deposition steps \cite{probst2024fabrication}. In this work, we examine one aspect of the increased fabrication complexity: undesired partially conformal layering, an artifact that arises in multilayer structures due to imperfect planarization between layer depositions (\figref{fig:all_layers}e), causing fabricated devices to deviate from their simulated counterparts and degrading device performance. Our previous work addressed this by limiting upper-layer deposition to regions unaffected by partially conformal layering \cite{probst2024fabrication}; while this approach has been shown to narrow the gap between experimental and simulated performance \cite{GFGC_conformal}, it also excludes the top layer from regions where its presence would benefit performance. A more versatile solution, therefore, is to develop a mapping function that predicts the resulting device geometry \textit{assuming} partially conformal layering from the set of design parameters. Then, the design parameters are optimized for maximal performance in the predicted geometry with partially conformal layering. This approach captures the full performance benefits of multilayer designs while faithfully representing the structure's real-world geometry, enabling devices that can reliably deliver the coupling efficiency and wavelength controllability that high-performance applications demand.

\subsection{Overview}
The paper is organized as follows: in \secref{sec:to}, we summarize previous efforts in TO and detail the specific optimization setup used in this work. In \secref{sec:vcsel}, we detail optimizing structures with a constant thickness conformal overlay (\secref{sec:conformal_mapping}) and numerically optimize a grating reflector atop a VCSEL cavity that maximizes differential modal reflectivity to maximize the single mode operation band (\secref{sec:vcsel_grating}). In \secref{sec:multilayer}, we examine unwanted partially conformal layer artifacts in multilayer stackups. First, we demonstrate the techniques necessary to optimize such structures using TO, highlighting the ways that the design phase can be tailored to the process-specific material stackups (\secref{sec:multilayer_mapping}). Then, we optimize a dual-layer silicon-on-insulator (SOI) grating coupler with significant partially conformal layering (\secref{sec:bilayer_gc}). Finally, we examine the effect that various depths of the conformal dip have on grating coupling performance, testing a grating coupler designed for conformal layering and one designed for planar layers (\secref{sec:conformal_sweep}).

\section{Topology Optimization} \label{sec:to}
In this section, we review the density-based photonic TO approach developed in our earlier work \cite{probst2024fabrication,hammond_constraints,hammond2022high}, which provides the basis for the advancements shown in this work. Photonic TO reformulates design as an optimization problem which minimizes $N$ objective functions, $(f_1,...,f_N)$, evaluated from Maxwell's equations at $M$ discrete frequencies, subject to $K$ constraint functions $(g_1,...,g_K)$ imposed on the geometry. The resulting problem is written as:
\begin{equation}\label{eq:obj}
    \begin{matrix}
         \min_{\vectrho}\left[\displaystyle\sum\limits_{n=1}^N\bar{g}_n\big(f_n(\vect{E}),\vect{q}_n\big)+\sum\limits_{k=1}^K\bar{g}_{N+k}\big(g_k(\vectrho),\vect{q}_{N+k}\big)\right] & n\in\left\{1,...,N\right\},\ k\in\{1,...,K\}\\
        \maxwell & m\in\left\{1,2,...,M\right\}\\
        0\leq\vectrho\leq1&\
    \end{matrix}
\end{equation}
where $\vect{E}$ and $\vect{J}$ are the electric field and source current respectively, and $\bar{g}_i,\ i\in\{1,...,N+K\}$ are differentiable spline-based scaling functions \cite{messac1996physical} used to weigh the figure of merit (FOM), $f_n$ and constraint functions, $g$, based on the user defined lists $\vect{q}$. The permittivity distribution $\varepsilon$ is generated from a set of latent (unfiltered) design variables $\vectrho$ via a mapping function; in this work, the three-field (filter-project) parameterization \cite{threefield,sigmund_overview} is used. The filter, $w(x)$, is a conic (or, in 1D, triangular) kernel applied to the latent variables to obtain filtered variables, $\widetilde{\rho}$:
\begin{equation}\label{eq:filter}
    \begin{matrix}
        \widetilde{\vectrho}=w(\vect{x})*\vectrho && w(\vect{x})=
    \begin{cases} 
      \frac{1}{a}\left(1-\frac{|\vect{x} - \vect{x}_0|}{R}\right) & \vect{x} \in \mathcal{N} \\
        \quad 0 & \vect{x} \notin \mathcal{N}
   \end{cases}
    \end{matrix},
\end{equation}
where $R$ sets the filter radius (and hence the minimum feature size) and $a$ normalizes the kernel. The filtered variables are subsequently pushed toward binary (0/1) values through a smoothed hyperbolic-tangent projection:
\begin{equation}\label{eq:threshold}
    \bar{\vectrho}=\frac{\tanh \left(\beta\eta\right)+\tanh \left(\beta\left(\vect{{\widetilde{\vectrho}}}-\eta\right)\right)}{\tanh\left(\beta\eta\right)+\tanh\left(\beta\left(1-\eta\right)\right)},
\end{equation}
with parameter $\beta$ controlling the sharpness of the projection and $\eta$ its center point. The voxels in the projected design parameters that correspond to a 1 in $\bar{\vectrho}$ are called ``solid'', and those corresponding to a 0 are called ``void''. Finally, the relative permittivity, $\vect{\varepsilon}_r$, is obtained by linearly interpolating the values of the void/solid projected design parameters onto the minimum and maximum dielectric permittivity.
\begin{equation}\label{eq:permittivity}
    \vect{\varepsilon}_r(\bar{\vectrho})=\vect{\varepsilon}_\text{min}+\bar\vectrho\big(\vect{\varepsilon}_\text{max}-\vect{\varepsilon}_\text{min}\big).
\end{equation}

Forward and adjoint electromagnetic simulations are carried out using the open-source finite-difference time-domain (FDTD) solver Meep \cite{meep}. Device performance metrics and their gradients with respect to every design variable, across all target wavelengths, are computed together via the hybrid time-frequency adjoint method, requiring only two full Maxwell solves regardless of the number of frequency points \cite{Strang07,steven_adjoint,hammond2022high}. These gradients are then propagated back through the three-field parameterization using a vector-Jacobian product (VJP) \cite{baydin2018automatic}, implemented with the automatic-differentiation package Autograd \cite{maclaurin2015autograd}. Finally, the latent design variables $\vectrho$ are updated using this gradient information within the globally convergent method of moving asymptotes (GCMMA) \cite{svanberg_mma}, as implemented in the open-source NLopt library \cite{nlopt}.

\section{Uniform Conformal Overlays}\label{sec:vcsel}

We implement a framework for inverse design of arbitrary aperiodic multilayer topologies with uniform conformal overlays, producing ternary structures where each pixel in the design region may take the value of the core material, cladding material, or overlay material, while requiring no additional design parameters over conventional binary topology optimized gratings of the same dimension. This framework is differentiable, compatible with our existing automatic differentiation codebase, and compatible with geometric design rule check (DRC) constraints. As a motivating example, we design a mode-selective semiconductor grating reflector with a SiN passivation layer for a VCSEL cavity enabling enhanced single mode operation. The uniform conformal overlay mapping is applied during the entire grating reflector optimization, yielding highly performant and inherently fabricable grating reflector geometries.

\subsection{Uniform Conformal Overlay Mapping Function}\label{sec:conformal_mapping}
\begin{figure}[htp]
    \centering
    \includegraphics{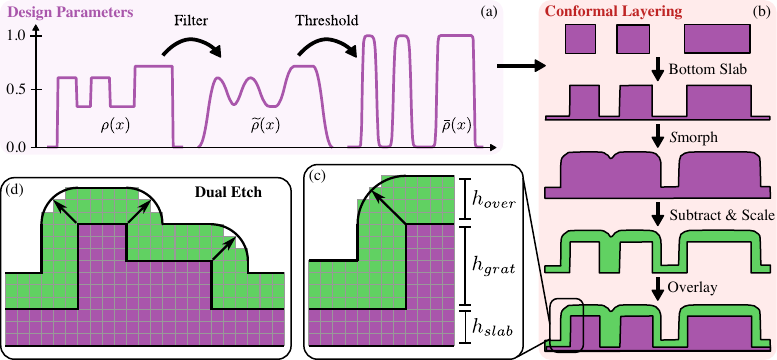}
    \caption{Single-etch and dual-etch parameters with a constant thickness conformal overlay. (a) The latent design parameters are projected using the filter-threshold approach. (b) The conformal overlay is accomplished by first vertically stacking the thresholded parameters and adding a bottom slab, then dilating to produce an outline of constant thickness. Permittivities are adjusted as needed, and the stacked parameters are added to the outline. (c) The final parameters contain a constant thickness overlay, where the dilation amount controls the thickness of the overlay. (d) Multilayer parameters without overhangs can be achieved with the same dilation process.}
    \label{fig:conformal_mapping}
\end{figure}

Enabling TO with conformal overlays requires a significant modification to the TO pipeline. Specifically, we augment the three-field scheme to include an additional conformal overlay step. To begin, the design parameters are filtered, projected, then vertically stacked, as in the conventional mapping from 1D design parameters to 2D permittivity voxels (\figref{fig:conformal_mapping}a). Next, a slab of solid phase with height $h_\text{slab}$ is appended to the bottom of the design array (\figref{fig:conformal_mapping}b). This provides a ``floor'' for the next step, a dilation of constant thickness, corresponding to the thickness of the conformal overlay. The slab is necessary to include overlay material in the gaps between solid phase, respecting the physical reality of a deposited passivation layer. The slab height is parameterized to map to the desired etch depth: if the material below the grating layer is the same material as the solid phase slab (e.g., for a partially etched semiconductor cap layer), the slab remains as the unetched portion of the layer; otherwise, it is subtracted. We use the 2D $\mathcal{S}$Morph operator \cite{kirszenberg2021going} for dilation — a differentiable operator that, unlike harmonic dilation (well-conditioned only for binary geometries \cite{svanberg2013density}), handles grayscale geometries directly and enables the inclusion of conformal overlay at the grayscale stage of the design process. The initial stacked 2D design parameters are then subtracted from the dilated parameters, leaving just the overlay (\figref{fig:conformal_mapping}b). The overlay parameters are then multiplied by a constant corresponding to the relative permittivity of the solid, overlay, and void phase materials: 
\begin{equation}
    m = (\varepsilon_{\text{Overlay}} - \varepsilon_{\text{Void}}) / (\varepsilon_{\text{Solid}} - \varepsilon_{\text{Void}}),
\end{equation} retaining the linear mapping from design parameter to permittivity voxels (\eqnref{eq:permittivity}).

This mapping has severable desirable properties. First, the $\mathcal{S}$Morph operator and array scalings and additions are differentiable operations that can act on both grayscale and binary structures, enabling their use in gradient-based optimizations \cite{hammond_constraints,zhou_geometric,probst2024fabrication,hammond2022high}. As the three-field parameterization is unchanged in this augmented mapping function, constraints such as minimum feature size, area, and enclosed area remain compatible. Also, this mapping function is easily extended into 3D design as the $\mathcal{S}$Morph function permits 3D dilation. Furthermore, the dilation kernel can be tuned to tune the conformal layer profile to match a real fabrication process. For example, the dilation kernel also need not be isotropic: an elliptical kernel sets the sidewall and top-surface overlay thicknesses independently, which effectively models deposition scenarios where the sidewall thickness differs from the flat regions (see Supplement 1, Section S1). Finally, the thickness of the conformal overlay can be varied in a robust optimization scheme to mitigate sensitivity to overlay thickness \cite{hammond_constraints}.

\subsection{Mode-Selective VCSEL Gratings}\label{sec:vcsel_grating}
\begin{figure}
    \centering
    \includegraphics{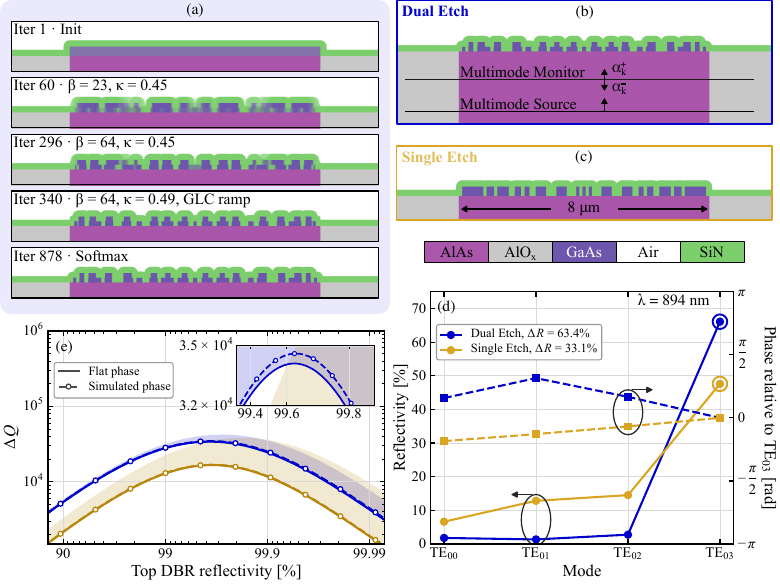}
    \caption{Optimization of the differential modal reflectivity $\Delta R$ of single-etch ($h_\text{grat} = 300 \ \text{nm}$) and dual-etch ($h_\text{grat} = [150\ \text{nm}, 150\ \text{nm}]$)  GaAs grating reflectors with SiN conformal overlay for maximal reflectivity in the TE$_{03}$ mode. (a) Selected optimization iterations for the dual-etch device showing ramping of thresholding parameter $\beta$ and layer projection parameter $\kappa$, as well as final softmax polishing. (b,c) The final $8\ \upmu$m-wide dual-etch and single-etch structures with conformal SiN overlay, shown with the source and monitor planes. (d) Forward pass simulation (with $\beta=\infty$) of power reflectivity (left) and modal phase (right) for the dual-etch (single-etch) structure, exhibiting 1.78, 1.34, 2.75, and 66.10\% (6.61, 12.83, 14.51, and 47.61\%) reflectivity for the first four TE modes respectively. The chosen mode, $\text{TE}_{03}$, is indicated by a concentric circle marker. (e) Assessment of viability for real DBR+grating mode selectivity via $Q$-factor cavity analysis. The bottom DBR reflectivity is fixed at 99.9\% and the top DBR reflectivity is swept. The solid line assumes flat DBR+grating modal phase response, dashed line the simulated phase response (d), and the shaded region shows all possible modal phase combinations. The top of the shaded region corresponds to the chosen mode in phase and all other modes perfectly out of phase.}
    \label{fig:vcsel_performance}
\end{figure}

We design mode-selective etched semiconductor grating reflectors with uniform conformal SiN passivation overlays to enhance the single-mode operation regime for high-aspect-ratio rectangular VCSELs. We accomplish this by maximizing the reflectivity of a chosen transverse mode and encouraging loss in the other modes, \eqnref{eq:modeselect}. This differential modal reflectivity, $\Delta R$, is directly related to the threshold gain of transverse modes in the cavity: a higher reflectivity indicates a lower threshold gain, therefore, a wider single-mode operation band can be realized by maximizing the $\Delta R$ (and thus differential modal gain) between the chosen mode and other cavity modes. To demonstrate this, we design single-etch ($h_\text{grat} = 300\text{ nm}$) and dual-etch ($h_\text{grat} = [150\text{ nm}, 150\text{ nm}]$) etched semiconductor grating reflectors with 200~nm uniform SiN overlay. The optimization is done in 2D (xz plane), using an effective index method that reduces the structure to 2D by assuming single-mode confinement into the page \cite{Khurana2026VCSEL}. The design region consists of a conventional oxidized AlAs aperture with a GaAs cap layer which is etched to form the grating reflector. A conformal SiN passivation layer is deposited onto the grating, a standard III-V fabrication process. For computational efficiency, we simplify to a single pass model, consisting of an oxidized aperture butt-coupled to the etched grating layer, by removing the conventional epitaxially grown distributed Bragg reflector (DBR) stacks and the multiple quantum wells because they are not inherently mode selective (they can be effectively modeled as slab indices since all transverse confinement is performed by the oxide aperture) \cite{coldren2012diode}. In this abstracted model, the transverse modes of the cavity are directly incident on the top side grating reflector \cite{Khurana2026VCSEL}. Thus, all mode selectivity is performed by a combination of the transverse oxidation aperture, which determines the total number of transverse modes that are confined in the cavity, and our optimized grating reflector, which performs mode selectivity according to the FOM:

\begin{equation}\label{eq:modeselect}
    f = \frac{1}{N}\left[-10\log_{10}R_i - \sum_{k \ne i}^N 10\log_{10}\left(1 - R_k\right)\right], \quad R_k \equiv \left|\frac{\alpha_k^-}{\alpha_k^+}\right|^2
\end{equation}
where $\alpha_k^{+,-}$ is the mode overlap integral with the TE$_{0k}$ transverse cavity mode in the $+z$ (up) or $-z$ (down) direction (\figref{fig:vcsel_performance}b). This FOM (in a minimization optimization) serves to preferentially maximize the reflectivity of a chosen mode, TE$_{0i}$, while encouraging loss in the other modes. This is accomplished by dividing the simulation into $N$ parallel simulations, launching one mode per parallel simulation, calculating each reflection coefficient per mode and the combined FOM, and averaging the gradients of each mode. We quantify the resulting mode selectivity by the differential modal reflectivity between the chosen mode and the most reflective undesired mode:
\begin{equation}\label{eq:deltaR}
    \Delta R \equiv R_i - \max_{k \ne i} R_k.
\end{equation}
In the following examples, we choose an 8 $\upmu$m $\times$ 3 $\upmu$m oxide aperture (reduced to a single dimension by the slab effective index into the page), which yields 3 additional higher order modes with $n_\text{eff}$ within a margin of 0.01 below the fundamental mode. Expansion of this effective index margin parameter would permit higher order modes, which is perfectly compatible with our simulation and optimization paradigm. We arbitrarily choose to optimize the grating reflectors to maximally reflect the TE$_{03}$ mode and minimally reflect the TE$_{00}$, TE$_{01}$, and TE$_{02}$ modes at $\lambda=894$~nm.

The grating reflector is initialized with a uniform gray 8~$\upmu$m wide design region including SiN overlay (\figref{fig:vcsel_performance}a) and optimized for maximal performance  according to \eqnref{eq:modeselect}. Once the grayscale design has achieved satisfactory performance, thresholding and minimum feature size constraints, in this case 100~nm geometric linewidth and linespace (GLC), are gradually enforced concurrently with the FOM \cite{hammond_constraints} such that the design still retains its performance. Importantly, the overlay is included from the first iteration. Unlike the minimum feature size constraints, which penalize regions of the design space and can therefore be introduced gradually, the 200~nm SiN overlay substantially alters the index contrast of the grating teeth and hence the  response; we found that a design optimized without the overlay is a poor starting point once the overlay is introduced. In contrast, the conformal dip in \secref{sec:bilayer_gc} is a small perturbation of the planar geometry and can be introduced late in the optimization. Additionally, the dual-etch device utilizes a layer projection, which restricts top layer solid phase parameters only to regions where the bottom layer is solid, to ensure fabricability of the multi-etch grating reflector without overhangs \cite{probst2024fabrication, probst2024topology}. The geometric constraints, the projection parameter for the layering $\kappa$, and the thresholding parameter $\beta$ are ramped up until the devices achieve a completely ternary structure (GaAs, SiN, air) (\figref{fig:vcsel_performance}b,c). This yields performant designs; however, to maximize performance we also apply a post-processing polish step by modifying the optimization function from a simple sum over the undesirable modes to a softmax, 

\begin{equation}\label{eq:softmax}
    f_\text{softmax} = -R_i + \tau \ln \left( \sum_{k \ne i}^N e^{R_k / \tau} \right)
\end{equation}
where $\tau$ is a temperature parameter. As $\tau \rightarrow 0$, the second term approaches $\max_{k \ne i} R_k$, so this objective places direct optimization gradient pressure on the ``next worst'' mode, i.e., the mode with reflectivity closest to the chosen mode, which sets $\Delta R$ (\eqnref{eq:deltaR}). In practice, this objective slightly impairs the reflectivity of the chosen mode while significantly suppressing the reflectivity of the undesirable modes, resulting in moderately improved $\Delta R$ (see Supplement 1, Section S2 for a thorough review of the optimization trajectory). After the above optimization pipeline including softmax polishing, we perform a single forward pass simulation with step-function $\beta=\infty$ projection. The single-etch grating reflector achieves $\Delta R=33.1\%$, with individual modal reflectivities of 6.61\%, 12.83\%, 14.51\%, and 47.61\% for the $\text{TE}_{00}, \text{TE}_{01}, \text{TE}_{02}$, and $\text{TE}_{03}$ modes respectively. The dual-etch grating reflector achieves $\Delta R=63.4\%$, with individual modal reflectivities of 1.78\%, 1.34\%, 2.75\%, and 66.10\% (\figref{fig:vcsel_performance}d). 

To assess the viability of the optimized conformal overlay grating reflectors for real single-mode VCSEL design, we compute the differential quality factor $\Delta Q \equiv Q_i - \max_{m \ne i} Q_m$ for the two designed grating reflectors, assuming a fixed bottom DBR reflectivity of 99.9\% and sweeping the top DBR reflectivity from 90\% to 99.99\% (\figref{fig:vcsel_performance}e). The mirror-limited modal $Q$ of the DBR cavity + grating reflector structure is:

\begin{equation}\label{eq:Q}
    Q_m = \frac{4\pi n_g L_\text{eff}}{\lambda \ln\left(1/\left(R_{\text{top}}^{(m)} R_{\text{b,DBR}}\right)\right)},\quad R_{\text{top}}^{(m)} \equiv \left| \frac{r_{\text{t,DBR}} + r_{\text{g}}^{(m)}}{1 + r_{\text{t,DBR}} r_{\text{g}}^{(m)}}\right|^2
\end{equation}

where $r_\text{g}^{(m)}$ is the amplitude reflection coefficient of the $m$-th mode from the optimized grating reflector ($|r_\text{g}^{(m)}|^2 = R_m$) \cite{coldren2012diode}, $n_g$ is the group index, and $L_\text{eff}$ is the effective cavity length. Using $n_g = 3.6$ and $L_\text{eff} = 1.5~\upmu$m, we obtain the solid curves in \figref{fig:vcsel_performance}e for the dual-etch and single-etch grating reflectors, which yield maximized $\Delta Q$ for a top DBR reflectivity around 99.6\%.

This model assumes that the modal phase responses of both the optimized grating reflector and the abstracted DBR are flat, which is not true in practice. To address the grating reflector modal response, we conduct a forward pass check on the modal phase response of the grating reflector (\figref{fig:vcsel_performance}d). The complex grating reflection coefficient then becomes $r_{\text{g}}^{(m)} \rightarrow r_{\text{g}}^{(m)}e^{i\varphi_m}$, with $\varphi_m$ the simulated grating reflection phase. Including the grating phase in the $\Delta Q$ calculation yields the dashed lines in \figref{fig:vcsel_performance}e. Because both of our objective functions are defined relative to the power reflectivity $R_k$, there is no phase information in the optimization. We observe the result is similar to that of the flat phase profile (\figref{fig:vcsel_performance}e inset). Regarding the DBR, the modal phase response for any designed DBR cavity can be computed from the phase velocity for each transverse mode that lives in the cavity, and included as the 0-phase reference for $\varphi_m$. The shaded region above the solid curves in \figref{fig:vcsel_performance}e corresponds to different combinations of grating reflector and DBR relative modal phase responses, with the top edge of the region (maximum $\Delta Q$) corresponding to the case where the chosen mode is perfectly in phase with the DBR and each of the undesirable modes is perfectly out of phase. By characterizing a designed DBR and including its phase response in the objective function with phase-sensitive topology optimization, or by co-designing a DBR together with a grating reflector, this method could in principle yield a mode-selective grating reflector that approaches this limit.

At the $\Delta Q$ maximum (reached with top DBR reflectivity of 99.6\%), the dual-etch grating reflector produces $\Delta Q = 3.5\times10^4$ using the simulated grating reflector phase, and the single-etch grating reflector yields $\Delta Q = 1.7\times10^4$. Rather than lowering the $Q$ of undesirable modes, the grating reflector serves to preferentially raise the $Q$ of the chosen mode. $Q_i$ increases from $1.5\times10^4$ (DBR alone) to $5.4\times10^4$ (DBR + dual-etch grating reflector) or $4.4\times10^4$ (DBR + single-etch grating reflector). The grating reflectors may be considered as relaxations of the DBR design requirement: a 99\% top DBR combined with the dual-etch grating reflector gives $Q_i = 3.7\times10^4$, comparable to a bare DBR with 99.9\% reflectivity. As the top DBR reflectivity approaches 99.99\%, the DBR reflectivity dominates, reducing mode selectivity. We note that $\Delta Q$ is simply a convenient method to quantify the grating reflectors, rather than a  laser design metric; the differential threshold gain, which scales with $1/Q_m - 1/Q_i$, increases as the top DBR reflectivity decreases, at the cost of a higher overall threshold. Additionally, the calculated $Q$ values are mirror-limited upper bounds assuming a lossless, mode-agnostic DBR model. Combined with the above phase discussion, these results indicate that the optimized grating reflectors presented here provide meaningful mode selectivity at practical DBR reflectivities, and serve as a starting point for DBR + grating reflector co-design of single mode VCSELs.

\section{Partially Conformal Multilayer Stackups}\label{sec:multilayer}
Multilayer fabrication provides greater optical controllability but incurs additional fabrication realities that must be captured in the design and simulation. In this section, we examine dual-layer structures where the bottom layer exhibits imperfect planarization, resulting in a partially conformal top layer (\figref{fig:all_layers}e). We demonstrate the functionality of the design technique with a dual-layer grating coupler.

\subsection{Partially Conformal Mapping Function} \label{sec:multilayer_mapping}
We develop a mapping function to generate dual-layer photonic structures with known partially conformal layering from the set of design variables. The mapping function detailed here assumes two sets of 1D design parameters mapping to a dual-layer 2D structure, but the mapping function is easily transferable to a 3D geometry.

\begin{figure}
    \centering
    \includegraphics{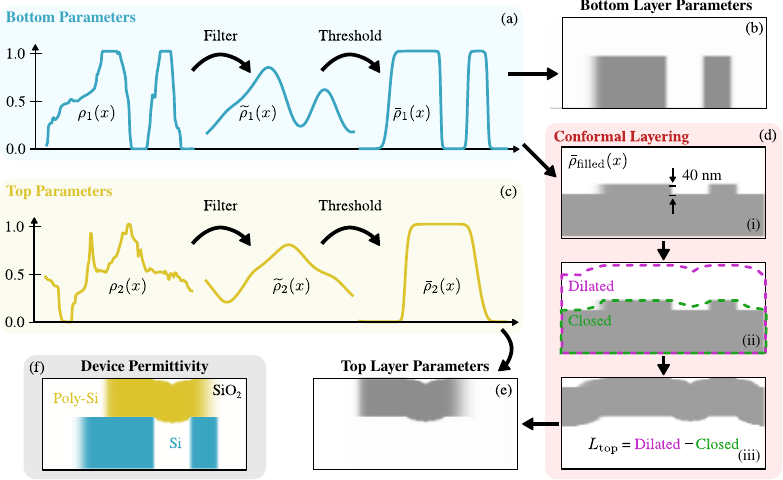}
    \caption{A TO-compatible mapping function for dual-layer devices where the top layer has partially conformal layering. (a,c) The bottom and top design parameters are projected using the filter-threshold approach. (b) The bottom design parameters are vertically extruded to the layer thickness. (d) To generate the nonplanar top layer, (i) the thresholded bottom parameters are vertically stacked above a block of solid phase. (ii) The closed bottom parameters are subtracted from the dilated bottom parameters, which produces (iii) the nonplanar top layer structure, $L_\text{top}$. (e) The thresholded top parameters mask $L_\text{top}$, (f) the two layers are added to form the device permittivity.}
    \label{fig:two_layer_mapping}
\end{figure}

To begin, the bottom latent design parameters, $\rho_1$, are mapped using the conventional three-field parameterization (\figref{fig:two_layer_mapping}a) and vertically stacked to the layer height (\figref{fig:two_layer_mapping}b). The top layer, because it is determined by the nonplanar surface of the bottom layer, is crafted from both the top and bottom parameters, with the surface topography (dips and bumps) coming from the bottom layer and the etch pattern from the top design parameters, $\rho_2$. To build the topography, we first construct a structure, $\bar{\rho}_\text{filled}$, consisting of two vertically stacked regions (\figref{fig:two_layer_mapping}d(i)). The lower region is a uniform slab of solid material. The upper region is the thresholded bottom design pattern extruded vertically, so it is solid where the bottom layer is unetched and void where it is etched. The thickness of the extruded region sets the maximum depth of the conformal dip in the top layer (40~nm in this case). The slab thickness is chosen so that the two regions together match the total thickness of the bottom layer. The top surface of $\bar{\rho}_\text{filled}$ therefore sits 40~nm lower above etched regions than above unetched ones, and this is the surface that the top layer conforms to. We then generate the top layer by dilating $\bar{\rho}_\text{filled}$ by the top-layer thickness, $t_2$, using the $\mathcal{S}$Morph transform \cite{kirszenberg2021going}:
\begin{equation}
    \text{Dilated}=\mathcal{S}\text{Morph}(\bar{\rho}_\text{filled},t_2)
\end{equation}
This produces the upper boundary of the top layer shown by the pink curve (\figref{fig:two_layer_mapping}d(ii)). The smooth lower boundary of the top layer, to emulate realistic fabrication processes, is computed by eroding the dilated structure by $t_2$, shown by the green curve (\figref{fig:two_layer_mapping}d(ii)). This operation is called a closing:
\begin{equation}
    \text{Closed}=\mathcal{S}\text{Morph}(\text{Dilated},-t_2).
\end{equation}
Lastly, the unpatterned top layer structure, $L_\text{top}$, is the difference between the dilated and closed geometries (\figref{fig:two_layer_mapping}d(iii)):
\begin{equation}
    L_\text{top}=\text{Dilated}-\text{Closed}.
\end{equation}
Physically, $L_\text{top}$ represents the structure of the top layer after deposition and before etching (i.e. a fixed thickness deposition on the nonplanar surface). The etch pattern (i.e. grating lines) in the top layer is determined from the projected parameters, $\bar{\rho}_2$ (\figref{fig:two_layer_mapping}c), acting as a mask on the nonplanar layer structure $L_\text{top}$ (\figref{fig:two_layer_mapping}e). The top and bottom layer parameters are added and then linearly interpolated onto the core and cladding permittivities, yielding the complete device permittivity (\figref{fig:two_layer_mapping}f).

The mapping is parameterized such that it can be tailored to an arbitrary process-specific conformal geometry. The maximum amount of the conformal dip in the simulation (40~nm in this example) is tuned by varying the amount of solid region and design parameters stacked to form $\bar{\rho}_\text{filled}$ (\figref{fig:two_layer_mapping}d(i)). We also note that if the solid slab region of $\bar{\rho}_\text{filled}$ is one pixel thick and the top layer is unetched, this mapping function is similar the mapping function in \secref{sec:conformal_mapping}, although the physical differences models (thin film vs dual-layer) make it useful to consider them separately and necessitate different optimization strategies. While the partially conformal mapping produces a realistic structure for the top layer, additional transforms could be used to shape the exact nonplanar structure to represent known cross-sectional profiles. Furthermore, this process is readily extensible to 3D geometries by using a 3D implementation of the erosion/dilation transforms \cite{probst2025integrated}. Finally, real conformal layering is a stochastic process variation which can modeled via robust optimization over a set of design parameters comprising the conventional planar mapping and the conformal mapping, yielding a device optimized for both partially conformal layering and perfectly planar layering \cite{hammond_constraints}.

\subsection{Dual-layer Grating Coupler Design} \label{sec:bilayer_gc}
\begin{figure}
    \centering
    \includegraphics{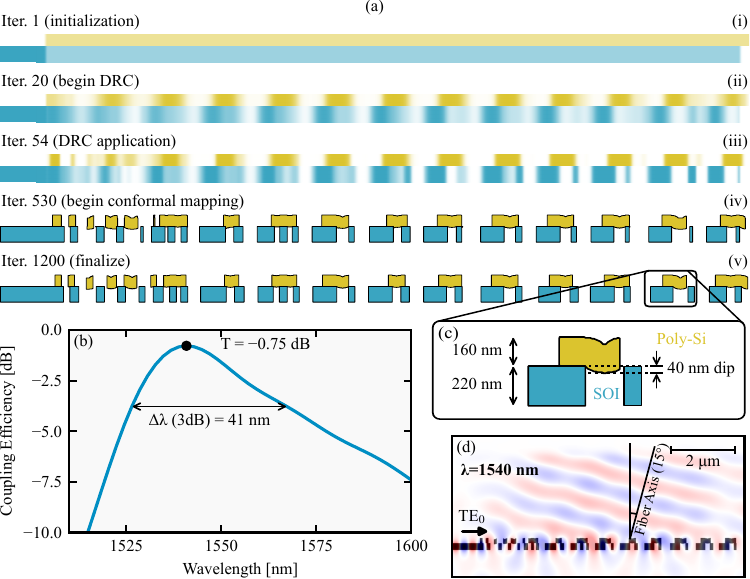}
    \caption{Dual-layer silicon grating coupler with an imperfectly planarized base layer causing partially conformal layering in the upper layer. (a) The grating is initialized with a grayscale geometry and optimized to maximize coupling to an SSMF at $\lambda=1540$~nm. (a,i-v) DRC constraints and the custom conformal mapping function are applied to achieve a fabricable design where (c) the 160~nm poly-Si top layer dips as much as 40~nm into a 220~nm Si base layer. (b) The grating achieves $-0.75$~dB coupling efficiency and 41~nm 3-dB bandwidth centered at the design wavelength $\lambda=1540$~nm and (d) emitting at a 15$^\circ$ incidence angle.}
    \label{fig:coupler_performance}
\end{figure}

We present the numerical optimization of a dual-layer Si grating coupler using a material platform with a partially conformal second layer dipping up to 40~nm into the lower layer. 1D grating couplers are highly sensitive to layer structure, making them an ideal device for demonstrating the conformal mapping function. The specifications of the optimization are as follows: the grating employs a 220~nm Si waveguide layer and a 160~nm poly-Si top layer which protrudes up to 40~nm down into gaps in the bottom layer (\figref{fig:coupler_performance}c). The grating is designed to couple light between an integrated waveguide and a standard single mode fiber (SSMF) positioned $1\ \upmu$m above the SOI with a launch angle of $15^\circ$. The design region is 10~$\upmu$m long and the minimum feature size is limited to 60~nm for both linewidth and linespacing. We define the FOM: 
\begin{equation}
    f= -10\log_{10}\left(\cfrac{|\alpha^+|^2}{P}\right),
\end{equation}
where $\alpha^+$ is the mode overlap with the forward propagating fiber mode and $P$ is the total power emitted by the source. The FOM is evaluated at $\lambda=$1540~nm. The optimizations were begun at a simulation resolution of $50\ \text{pixels}/\upmu$m and finished at an upscaled resolution of $100\ \text{pixels}/\upmu$m. The design-grid resolution was upscaled to twice the simulation resolution in order to accurately reconstruct the curved structure generated by the partially conformal layering \cite{hammond2022high}. Final simulations use further upscaled simulation resolution of $200\ \text{pixels}/\upmu$m.

The optimization is initialized with grayscale design parameters ($\rho=0.5$) on both layers (\figref{fig:coupler_performance}a(i)). We design the optimization in three stages. First, we assume perfectly planar layering and do not enforce any binarization or minimum size constraints (\figref{fig:coupler_performance}a(ii)). Next, we simultaneously optimize the FOM and the DRC constraints \cite{zhou_geometric} (\figref{fig:coupler_performance}a(iii)) while still assuming perfectly vertical layers. Finally, we use the conformal layer mapping function, continuing to optimize both the FOM and the DRC constraints (\figref{fig:coupler_performance}a(iv)). We find that devices with the highest performance are produced by optimizing first without constraints assuming perfectly vertical layers until the device reaches a highly performant design space and then applying the DRC constraints and the conformal mapping. Knowing when to activate the minimum feature size constraints and the conformal layer mapping requires skill by the designer and is aided by running parallel optimizations varying how the DRC constraints and conformal mapping are applied (see Supplement 1, Section S3 for the optimization trajectory).

This grating achieves $-0.75$~dB coupling efficiency centered at 1540~nm and 41~nm 3-dB bandwidth (\figref{fig:coupler_performance}b), which is very low-loss coupling for a dual-layer design without a bottom reflector (\figref{fig:coupler_performance}c), and nearly equivalent to our previous results assuming perfectly vertical layering \cite{probst2024fabrication}. The continuous wave (CW) excitation of the grating from the waveguide at $\lambda=1540$~nm shows that the majority of the power is directed upward at the intended $15^\circ$ launch angle (\figref{fig:coupler_performance}d). These results indicate that even significant unwanted partially conformal layering can be managed by the optimizer \cite{Lee2024PolarizationSplitterRotator} if it is accounted for during the design phase, and are promising results for reducing simulation-experiment performance discrepancies.

\subsection{Partially Conformal Layer Sweep}\label{sec:conformal_sweep}
We compare the performance of two grating couplers, each simulated across a range of conformal layering depths. The first is the grating designed for a 40~nm conformal dip in the upper layer (\secref{sec:bilayer_gc}), and the second is a grating designed under the assumption of perfectly planar layers. For a fair comparison between the two design methods, which we refer to as the partially conformal and planar methods, the resulting designs should differ only in the aspects specific to the conformal layering. Conveniently, the partially conformal optimization treats the layers as planar during its early stages before the conformal mapping is introduced. To produce the planar design, we therefore took the last iteration of the partially conformal optimization in which the layers were still assumed planar and continued optimizing from that point with planar layers throughout. Because both gratings share an identical design history up to the point where the conformal mapping is applied, any difference in their performance can be attributed directly to the effect of that mapping.

\begin{figure}[htpb]
    \centering
    \includegraphics{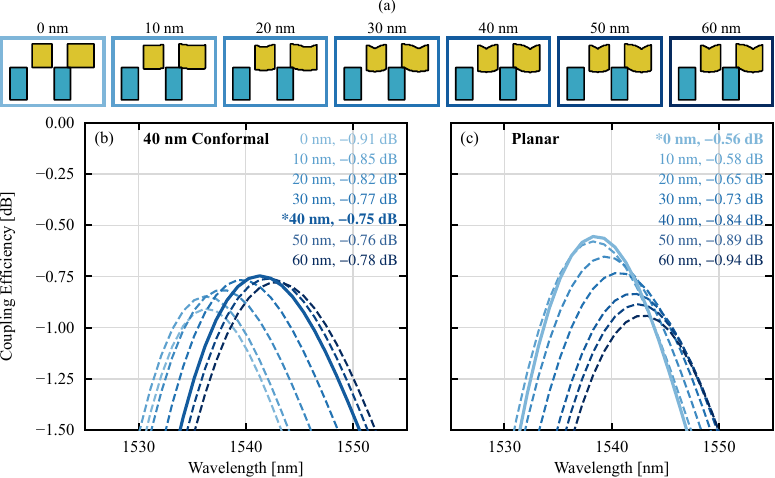}
    \caption{Grating couplers designed for 40~nm conformal dip and 0~nm conformal dip both simulated from 0~nm to 60~nm conformal dip in steps of 10~nm. (a) A subset of the parameters are shown at the simulated conformal dip depths. (b) The grating designed for 40~nm partially conformal layering has peak performance correctly centered at $\lambda=1540$~nm and peak performance when the conformal dip is 40~nm. The grating designed for perfectly vertical layers similarly has the peak performance centered at $\lambda=1540$~nm, but does experience performance increase with 10~nm conformal dip. Both gratings show $\leq0.1$~dB sensitivity in peak coupling efficiency when the amount of conformal dip is within 20~nm of the designed value.}
    \label{fig:coupler_sweep}
\end{figure}

The two grating couplers are simulated from 0~nm (i.e. planar layers) to 60~nm of conformal dip in steps of 10~nm (\figref{fig:coupler_sweep}a). The data show that coupling efficiency is maximized when both designs are simulated at the intended amount of conformal dip ($-0.75$~dB for the 40~nm dip conformal grating and $-0.56$~dB for the grating with planar layers), and a general blue-shift for shallower partially conformal layers (\figref{fig:coupler_sweep}b,c). Furthermore, both designs show $<0.1$~dB of degradation in peak coupling efficiency if the maximal conformal dip is within 20~nm of the designed amount. Finally, at the 40~nm conformal dip expected from the fabrication process used in this paper, the grating designed with the conformal mapping achieves $-0.75$~dB coupling efficiency, compared with $-0.84$~dB for the grating designed for planar layers, suggesting that incorporating the conformal dip into the optimization raises the performance ceiling under conformal fabrication reality.

\section{Conclusion}
This work enables the optimization of photonic structures with nonplanar conformal overlays and improves design adherence to fabrication realities by directly modeling these effects during the optimization. Incorporating these fabrication realities, rather than treating them as post-hoc fabrication error or restricting the design space to structures without conformal features, produces designs that closely resemble the fabricated structures and enable the optimizer to traverse a larger portion of the search space. We designed (1) a uniform conformal overlay mapping, which effectively models passivation depositions above a patterned surface, and (2) a partially conformal mapping function, modeling imperfect planarization in a fabrication process with multiple deposition steps. Both of these enhancements are readily included into a conventional TO pipeline and enhance the generalizability and fabrication awareness of TO algorithms. To illustrate the workflow using these enhancements, we designed two classes of optical gratings. First, mode-selective GaAs grating reflectors with a conformal SiN passivation layer achieved differential modal reflectivities of 33.1\% (single-etch) and 63.4\% (dual-etch) for the TE$_{03}$ mode of a high-aspect-ratio VCSEL aperture. Second, a dual-layer grating coupler with a partially conformal upper layer dipping up to 40~nm into the imperfectly planarized lower layer achieved $-0.75$~dB coupling efficiency to an SSMF and 41~nm 3-dB bandwidth. This grating outperforms a grating optimized assuming planar layers when both are evaluated at the 40~nm conformal dip. These case studies demonstrate the value of modeling the conformal layering during the design process.

Future work includes performing robust optimizations on conformal overlay parameters, including the layer thickness and magnitude of the conformality. The projection and transform methods described in this work are extensible to anisotropic conformal overlays (Supplement 1, Section S1), non-square etch profiles including blazed and angled sidewalls \cite{probst2025integrated}, and arbitrary conformal geometries which may be extracted from cross-sectional imaging of fabricated devices. Together with conventional geometric design rule constraints, these methods provide a key advancement toward first-time right photonic fabrication.

\begin{backmatter}
\bmsection{Funding} National Science Foundation (2052808).

\bmsection{Acknowledgments} This research was supported in part through research cyberinfrastructure resources and services provided by the Partnership for an Advanced Computing Environment (PACE) at the Georgia Institute of Technology, Atlanta, Georgia, USA.

\bmsection{Disclosures} The authors declare no conflicts of interest. The authors used Claude (Opus 5.5) solely for refinement of text and development of simulation and analysis scripts. AI was not used to generate scientific claims, references, data, or conclusions. The authors reviewed and take responsibility for all content.

\bmsection{Data availability} Data underlying the results presented in this paper are not publicly available at this time but may be obtained from the authors upon reasonable request.

\bmsection{Supplemental document} See Supplement 1 for supporting content.
\end{backmatter}


\bibliography{references}

\end{document}